**Original Paper**

# Sustaining model performance for covid-19 detection from dynamic audio data: Development and evaluation of a comprehensive drift-adaptive framework

Theofanis Ganitidis[1]; Maria Athanasiou[1], PhD; Konstantinos Mitsis[1], PhD; Konstantia Zarkogianni[2], PhD; Konstantina S. Nikita[1,3], PhD, MD

[1]School of Electrical and Computer Engineering, National Technical University of Athens, Athens, Greece

[2]Maastricht University, Faculty of Science and Engineering, Maastricht, The Netherlands

[3]University of Southern California, Viterbi School of Engineering, Los Angeles, USA

**Corresponding author:**
Theofanis Ganitidis
School of Electrical and Computer Engineering
National Technical University of Athens
9, Iroon Polytechniou Str.,
15780 Zografos, Athens Greece
Tel: 0030 - 210 - 7722285
Email: theogani@biosim.ntua.gr

**Abstract**

**Background:** The COVID-19 pandemic has highlighted the need for robust and adaptable diagnostic tools capable of detecting the disease from diverse and continuously evolving data sources. Machine learning models, particularly convolutional neural networks (CNNs), have shown promise in this regard. However, the dynamic nature of real-world data can lead to model drift, where the model's performance degrades over time as the underlying data distribution changes. Addressing this challenge is crucial to maintaining the accuracy and reliability of these models in ongoing diagnostic applications.

**Objective:** This study aims to develop a comprehensive framework that not only monitors model drift over time but also employs adaptation mechanisms to mitigate performance fluctuations in COVID-19 detection models trained on dynamic audio data.

**Methods:** Two crowd-sourced COVID-19 audio datasets, the COVID-19 Sounds and COSWARA datasets, were used for development and evaluation purposes. Each dataset was divided into two distinct periods: development and post-development. A baseline CNN model was initially trained and evaluated using data (i.e., cough recordings) from the development period. To detect changes in data distributions and the model's performance between these periods, the maximum mean discrepancy (MMD) distance was employed. Upon detecting significant drift, a retraining procedure was triggered to update the baseline model. The study explored two model adaptation approaches: unsupervised domain adaptation (UDA) and active learning (AL), both of which were comparatively assessed.
**Results:** The application of the UDA approach led to performance improvement in terms of the balanced accuracy by up to 22% and 24% for the COVID-19 Sounds and COSWARA datasets, respectively. The AL approach yielded even greater improvement, corresponding to a balanced accuracy increase of up to 30% and 60% for the two datasets, respectively.
**Conclusions:** The proposed framework successfully addresses the challenge of model drift in COVID-19 detection by enabling continuous adaptation to evolving data distributions. This approach ensures sustained model performance over time, contributing to the development of robust and adaptable diagnostic tools for COVID-19 and potentially other infectious diseases.

## Introduction

The rapid spread of the novel coronavirus (SARS-CoV-2) and its associated disease, covid-19, has brought about a pressing need for accurate and timely diagnostic tools. Traditional diagnostic methods such as polymerase chain reaction (PCR) tests, while reliable, often involve invasive procedures and can be time-consuming. Consequently, there is a growing interest in developing additional diagnostic approaches that are non-invasive, affordable, scalable, and capable of delivering swift results[1].

Deep learning models have demonstrated exceptional capabilities across various domains, including medical diagnostics [2–8] and epidemiological surveillance [9]. Studies [10–13] have illuminated the potential of harnessing deep learning techniques for analyzing diverse data sources, such as clinical and biological biomarkers, CT scan imagery, and clinical characteristics, to predict the severity and progression of covid-19. In recent studies, the analysis of cough sounds has shown potential as a non-invasive modality for covid-19 detection [14–16]. Leveraging the power of deep learning, these models can extract crucial information from acoustic characteristics, aiding in the early identification of infected individuals.

However, in practice, the performance of deep learning models tends to decline during deployment and shows further deterioration over time. This phenomenon, commonly known as model degradation, can be attributed to various factors [17]. One contributing factor is the limited representation of the training data, which fails to capture the complexity of the problem space adequately. As a consequence, the model exhibits unexpected behavior when confronted with input samples lying outside the distribution of training examples [18,19]. Another significant factor is the dynamic nature of the system's environment, which undergoes continuous changes over time [18], posing challenges for a single model to maintain accurate predictions consistently. This factor is particularly critical in the context of covid-19, given the rapid and unpredictable changes due to several reasons, including the emergence of new viral strains.

The literature refers to these two factors as concept drift, which is the phenomenon where the input data and their relationship to the labels undergo changes over time.

Numerous attempts have been made in the past decade to precisely define concept drift [17,20–23]. In this paper, the definition from [22] is adopted, which states that concept drift occurs when either the data distribution changes or the underlying relationship between the input and output changes, or both.

Researchers have recognized the importance of addressing these challenges and have focused on learning in non-stationary environments [24] and mitigating the impact of concept drift [25–28]. Research studies have stressed the importance of integrating a model degradation detector within the learning framework [29] that assesses and tracks the system's performance after deployment to effectively manage the degradation in prediction accuracy. The level of degradation in the model performance serves as an indicator for detecting concept drift within the system. By incorporating these detection components, deep learning systems gain resilience against environmental changes, thereby mitigating the performance degradation of predictive models in this ever-changing setting.

Since the presence of concept drift between training data and real post-development data impedes the performance of deep learning models on out-of-distribution samples [25], applying the model on new data may necessitate adaptation. Automatic methods have emerged to tackle these challenges; however, collecting large-scale labeled datasets for different populations, emerging virus variants or new pandemics is an arduous task. When working with limited data, it is often necessary to employ more cost-efficient deep learning methodologies, such as unsupervised domain adaptation (UDA) and active learning (AL).

Domain adaptation is a technique utilized when training and testing data come from different distributions [30] towards addressing the limited generalization abilities of predictive models. The goal is to adapt a model trained on a source domain to perform well on a target domain. This involves minimizing the distribution gap between the domains through learning domain-invariant features [31], weighing samples based on similarities [32], or using model-based techniques like domain adversarial networks [33]. These approaches can improve model generalization in real-world scenarios with varying data distributions as they enable learning from labeled data in the development set, which refers to the past, and applying this knowledge to solve tasks on post-development unlabeled data.

Active learning, on the other hand, is a machine learning approach where informative samples from a large unlabeled dataset are selected and labeled iteratively to train a model. The objective is to minimize the amount of labeled data needed while maximizing the model's performance [34,35]. A query strategy is selected to determine which unlabeled samples should be labeled. Various strategies exist, such as uncertainty sampling [36,37] or diversity sampling [38]. Given an initially trained model, the chosen query strategy is then applied to the unlabeled dataset, identifying the most informative samples based on the selected criterion. These selected samples are labeled either manually by domain experts or through an automated process. The newly labeled samples are incorporated into the labeled dataset, which is used for retraining the model.

The majority of recent studies focusing on covid-19 detection based on the use of audio recordings have primarily employed supervised learning techniques [15,16,39–42], whereas the exploration of methods relying on concept drift detection, model degradation detection, UDA, and AL has been limited [27,43]. Nonetheless, UDA and AL approaches appear to be highly promising and well-suited for addressing new pandemics, as they enable the development of reliable models even when confronted with the emergence of novel variants.

In this paper, a comprehensive framework is introduced for the diagnosis of infectious diseases, focusing on covid-19 detection from cough sounds. The framework leverages deep learning models combined with supervised and unsupervised learning methodologies to monitor and mitigate model degradation and concept drift. The development and evaluation of the proposed framework is demonstrated on covid-19 data due to their continuously evolving epidemiological and virological characteristics, arising from the complex interplay among the virus, humans, vaccines, and environments. The maximum mean discrepancy (MMD) [44] distance is firstly employed as a metric to quantify the dissimilarity between temporal data distributions. By monitoring the MMD distance between batches of post-development data and data from the initial development period, the framework detects changes in both the data and the model's performance while also providing insights into the impact of the pandemic's evolution on the trained models' diagnostic accuracy. In case concept drift is detected, a retraining process is initiated, including two adaptation methods: i) a UDA process which leverages labeled development data along with unlabeled post-development data to align their distributions and adapt the model to novel data instances and ii) an AL strategy, aiming at selecting informative data to include them with their labels in the retraining process. To the best of the authors' knowledge, this is the first work leveraging UDA and AL approaches towards mitigating the impact of evolving data dynamics on model performance in the case of covid-19 with the ultimate goal to enhance reliability in covid-19 detection and potentially across various diverse epidemiological contexts.

## Methods

### Datasets

The COVID-19 Sounds dataset is a collection of respiratory sound recordings associated with covid-19 infections, which were acquired through a crowd-sourcing platform launched in April 2020. It includes demographic characteristics (i.e., age, gender) along with participant-reported information about medical history and symptoms. It also comprises audio clips of voluntary cough, breathing, and speech captured from healthy individuals and individuals with covid-19. A total of 36,364 participants contributed 75,201 samples to the project. Quality checks were performed on the audio samples to filter out incomplete or noisy recordings [16]. The data were collected in multiple languages, but for the present study the part of the dataset acquired from English-speaking participants [16] was solely considered to avoid language bias, corresponding to a total of 1461 samples, as shown in Table 1.

The COSWARA dataset is another crowd-sourced database, recorded between April-2020 and February-2022, which consists of various kinds of sounds, such as shallow and deep breaths, shallow and heavy coughs, sustained vowel phonation (i.e. /ey/ as in made, /i/ as in beet, and /u:/ as in cool), and number counting from one to twenty (normal and fast-paced). Alongside this, information on the participants' covid-19 infection status, symptoms, co-morbidities (if any), gender, age, and broad geographical location is included. In the present study, 2746 shallow cough recordings were considered for composing the models' input space. Specific criteria were applied towards the exclusion of samples that had any of the following characteristics: missing, corrupted, or silent shallow cough recordings. This procedure resulted in 1996 samples, as shown in Table 1.

*Table 1: Partition of the used datasets into development and post-development sets.*

|  | #samples | |
|---|---|---|
|  | COVID-19 Sounds | COSWARA |
| Development set (covid-19-positive) | 1040 (452) | 1395 (165) |
| Post-development set (covid-19-positive) | 421 (270) | 601(482) |

Regarding data preprocessing, the setting described in [16] was applied. The audio signals were normalized, and leading and trailing silence was removed. Mel-spectrograms were calculated using a 25 *ms* window size, a 10 *ms* window hop, and a total of 64 Mel bins, encompassing frequencies ranging from 125 Hz to 7500 Hz. To handle the varying size of the Mel-spectrograms' time axis, the 0.9 quantile across all spectrograms was calculated. Subsequently, the spectrograms were either cropped or padded with repeated sections of the spectrogram accordingly. Finally, a sliding window approach was employed to extract segments from the spectrogram. The width of the window used was 0.96 *s*, while the window stride length was equal to half of the window's width (0.48 *s*). This setting resulted in a Mel spectrogram segment of a 64 × 96 size.

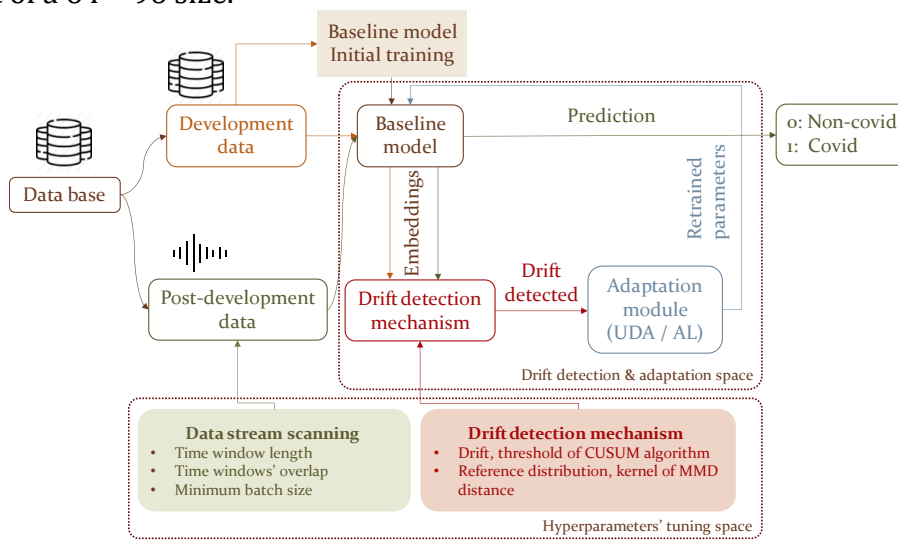

*Figure 1: Overview of the proposed framework. The data are divided in two sets, the labeled development set which is used for the initial training of the baseline model, and the unlabeled post-development set which is used to evaluate the proposed framework. The framework comprises three key modules: (i) The baseline model, initially trained on the development dataset for binary classification. (ii) The drift detection mechanism, which monitors the performance of the baseline model during the post-development period to detect data drifts. (iii) The adaptation module, triggered upon data drift detection, which utilizes either Unsupervised Domain Adaptation (UDA) or Active Learning (AL) to retrain the model for improving performance.*

To facilitate model training, the COVID-19 Sounds and COSWARA datasets were partitioned based on chronological order into a development set and a post-development set by applying a 70:30 ratio. The development set was further divided into training, validation, and testing subsets using a 60:20:20 split ratio, respectively. This division ensured that the model was trained on a representative portion of the data and validated and tested on separate subsets, promoting robustness and generalization.

**Proposed methodological framework**

The abstract architecture of the proposed framework is depicted in Figure 1. It comprises three distinct modules, combining a deep neural network with a drift detection mechanism and appropriate adaptation modules with the aim to address differences between data distributions of the development and post-development periods. These modules are explained below:

- Baseline model: A baseline model based on Convolutional Neural Networks (CNN) processes input data instances and estimates the probability of covid-19 presence.
- Drift detection mechanism: This module is responsible for the identification of shifts in the data, implying changes in covid-19 detection patterns. It monitors the performance of the baseline model through the detection of significant discrepancies between the development data and the post-development data. To this end, a modified version of the CUSUM (CUmulative SUM) algorithm is utilized, employing the MMD distance for measuring the distance between data distributions from the development and post-development periods. A set of hyperparameters, which are appropriately adjusted, are included in the CUSUM algorithm (i.e., drift, threshold) and the MMD distance (i.e., reference distribution, kernel).
- Adaptation module: The adaptation module facilitates the adaptation process within the system. It enables the model to dynamically adjust and learn from post-development data, ensuring continuous improvement and robustness against evolving covid-19 characteristics. Two different approaches employing divergence-based UDA and AL are investigated for harnessing post-development, unlabeled data towards model retraining with the aim of enhancing the performance and improving the generalization abilities of the baseline model.

The proposed framework's operation is based on the adoption of a batch-based approach for the processing of data instances and the application of a fixed time window (with parametrized overlap between successive windows) to monitor the data stream for changes. The time window length, overlap between successive time windows, and minimum batch size along with the hyperparameters of the drift detection mechanism are appropriately validated to ensure optimal performance for each dataset.

The development and evaluation of the proposed framework was based on the use of cough recordings from the COVID-19 Sounds [39] and COSWARA [45] datasets. Both datasets were partitioned into development and post-development sets based on chronological order. Figure 2 illustrates the partition of data, while Table 1 summarizes the distribution of covid-19 positive samples for both datasets.

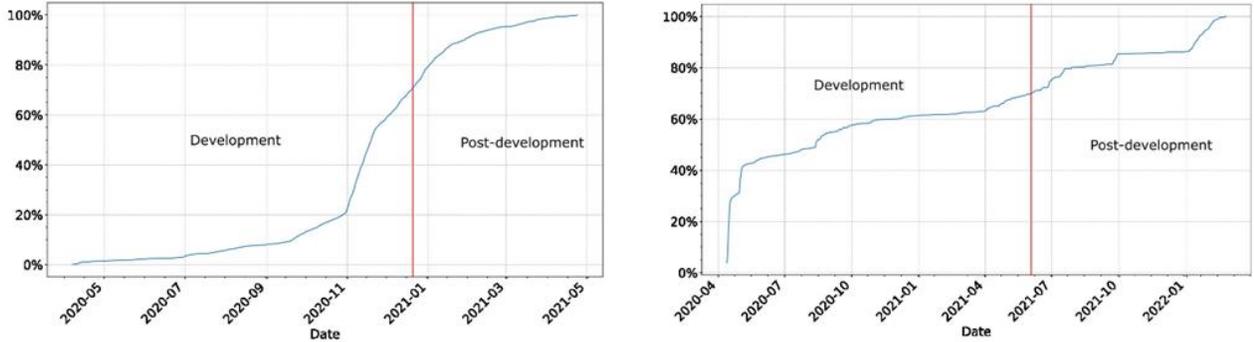

*Figure 2: COVID-19 Sounds (left panel) and COSWARA (right panel) data streams over time. A 70:30 partition of the data stream into development and post-development sets was applied. Red line indicates the division between development and post-development data. The development set (further partitioned into training, validation, and test subsets using a 60:20:20 ratio) was used for developing the baseline model. Special care was taken to ensure that no overlap existed in terms of data from the same participant among the training, validation, and test subsets, as well as between the development and post-development periods.*

*Baseline model*

The baseline model of the proposed framework was built upon the widely used VGGish pretrained model [46,47] which was selected due to its remarkable performance on audio classification tasks [16,48]. The VGGish model is a deep CNN trained on a large-scale audio dataset to learn hierarchical representations of audio signals. In the framework of the present study, the VGGish model was utilized to extract discriminative features from segments of Mel spectrograms with the aim of capturing relevant acoustic patterns and distinguishing covid-19 coughs from non-covid-19 coughs. Figure 3 shows the general architecture of the used baseline model.

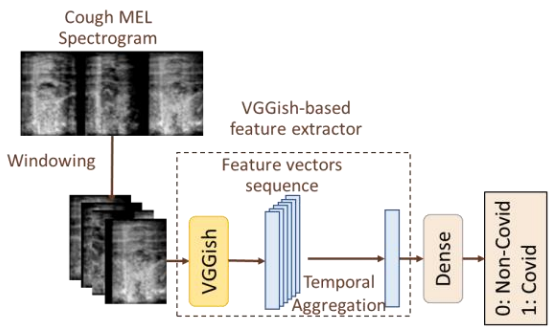
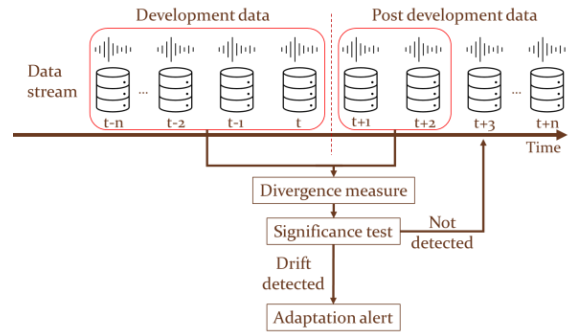

*Figure 3: Baseline model architecture.*   *Figure 4: Proposed drift detection mechanism.*

To adapt the VGGish model to the specific task of the present study, a time-distributed approach was employed. To this end, the VGGish feature extractor was applied on each segment of the Mel spectrogram, resulting in a sequence of feature vectors that represented the temporal evolution of acoustic characteristics within the cough signal. In order to summarize the temporal dynamics captured by the model, the mean value for each feature across the entire sequence was calculated. Following the temporal aggregation, a dense layer with a single node based on the non-linear sigmoid activation function was employed to process the aggregated feature vectors and calculate the final output of the model.

In order to address the imbalanced nature of the datasets, the binary focal cross-entropy loss function was employed for training the baseline model due to its ability to focus on rare examples [49]. This loss function effectively assigns higher weights to misclassified samples, thereby alleviating the impact of class imbalance, and improving overall performance. For optimization, the Adam (Adaptive Moment Estimation) optimizer was employed due to its efficient and adaptive nature [50]. A learning rate equal to $10^{-4}$ was used, while the exponential decay rate for the first and second moment estimates was 0.9 and 0.999, respectively.

During training, the labeled samples of the development set were considered for minimizing the chosen loss function. A batch size of 32 and 100 epochs were used, which is a commonly used default training scheme employed in multiple studies [51,52]. The validation score was employed for monitoring the model's convergence and an early stopping regularization technique was applied. After convergence, the performance of the trained model was evaluated on a test subset sampled from the development period and on the entire post-development set.

## *Drift detection mechanism*

The proposed drift detection mechanism entailed divergence monitoring using the MMD distance and the implementation of the CUSUM algorithm [53,54] for generating drift alerts. The data were utilized in a chronological order towards monitoring the performance of the model. A batch-based approach was adopted for monitoring and processing data instances. An overview of the proposed drift detection mechanism is provided in Figure 4.

### Monitoring divergence with MMD distance

To effectively track the dissimilarity between the development data and post-development data, the MMD distance was adopted, which was computed by comparing the corresponding embeddings extracted by the VGGish feature extractor, as described in the previous subsection. These embeddings served as a representation of the data distribution and were used for calculating the MMD distances between batches of the post-development data and the development data, with the embeddings of the latter constituting the reference distribution. The MMD distance value between two batches of data is given by the following equation:

$$\begin{aligned} MMD(X,Y) &= \left\| \frac{1}{n_x} \sum_{i=1}^{n_x} \varphi(x_i) - \frac{1}{n_y} \sum_{i=1}^{n_y} \varphi(y_i) \right\|_H^2 \\ &= \langle \frac{1}{n_x^2} \sum_{i,j} k(x_i, x_j), \frac{1}{n_y^2} \sum_{i,j} k(y_i, y_j) \rangle_H \\ &= \frac{1}{n_x^2} \sum_{i,j=1}^{n_x} k(x_i, x_j) + \frac{1}{n_y^2} \sum_{i,j=1}^{n_y} k(y_i, y_j) \\ &\quad - \frac{2}{n_x n_y} \sum_{i,j=1}^{n_x, n_y} k(x_i, y_j) \end{aligned} \quad (1)$$

where:

- *X* represents the distribution of the embeddings of the development data $x_i$ (reference distribution), with $n_x$ embeddings in total.
- *Y* represents the distribution of the embeddings of post-development data $y_i$, with $n_y$ embeddings in total.
- $\varphi(.)$ represents the feature mapping function used to transform the embeddings into a high-dimensional space (VGGish model).
- $k(.,.)$ is a kernel function that computes the similarity between two inputs.
- $\langle .,. \rangle_H$ denotes the inner product in the Hilbert space induced by the kernel function.

In this work, the use of three different kernels (linear, polynomial of degree two, and Gaussian) was investigated.

### CUSUM algorithm

After calculating the divergence between the development and post-development data, an implementation of the CUSUM algorithm was deployed for detecting points of significant increase in the divergence measure. CUSUM is a change detection algorithm which is widely used to identify shifts or changes in time series data [55–57], particularly when the exact nature of the change is unknown or when there is a need to continuously monitor data for detecting changes. CUSUM is widely adopted for real-time monitoring and surveillance applications in various fields, including quality control, signal processing, and anomaly detection.

In this study, the CUSUM algorithm was tailored to match the specific characteristics of the deep learning model and the monitored MMD distance. The proposed implementation introduced the calculation of relative differences between successive values instead of their corresponding absolute values, thus enabling the original CUSUM algorithm to effectively align with the behavior of the MMD distance and the desired level of sensitivity to changes. Therefore, the drift and threshold values represented the tolerance range of relative change in successive values and the minimum cumulative relative change required to trigger a change detection event, respectively.

### *Adaptation mechanism*

Upon the triggering of an alert by the drift detection mechanism, an adaptation mechanism based on model retraining was activated to update the baseline model. The proposed adaptation mechanism aimed at enhancing the performance and improving the generalization abilities of the baseline model. Two different approaches based on UDA and AL were explored for the development of the adaptation mechanism.

### Unsupervised domain adaptation

The UDA approach involved feeding the model with a batch of post-development data samples, along with a batch of samples from the development set. The model was then trained jointly on two tasks: (i) to correctly classify the labeled development data, and (ii) to minimize the MMD distance between the embeddings of the development and post-development batches. In this way, the model was trained to solve task (i) using domain-invariant features (development and post-development data), aiming at the minimization of two loss functions. The first loss function considered the model's output on samples of the development dataset, essentially employing supervised learning. The second loss function was based on the divergence between the distributions of the post-development data and development data batches using the MMD distance. The Gaussian kernel:

$$k(x,y) = \exp\left(-\frac{\|x-y\|^2}{2\sigma^2}\right) \quad (2)$$

was selected to be used for the MMD distance calculation due to its ability to distinguish between distributions with differences in any order of moments [33,58] as demonstrated by its Maclaurin series representation:

$$k(x,y) = \exp\left(-\frac{\|x-y\|^2}{2\sigma^2}\right)$$
$$= 1 - \frac{\|x-y\|^2}{2\sigma^2} + \frac{\left(\frac{\|x-y\|^2}{2\sigma^2}\right)^2}{2!} - \frac{\left(\frac{\|x-y\|^2}{2\sigma^2}\right)^3}{3!} \quad (3)$$
$$+ \frac{\left(\frac{\|x-y\|^2}{2\sigma^2}\right)^4}{4!} - \cdots$$

In contrast, the linear kernel cannot distinguish between distributions with the same mean but different higher-order moments while the polynomial kernel of degree two is unable to differentiate between distributions that have the same mean and variance but differ in higher-order moments.

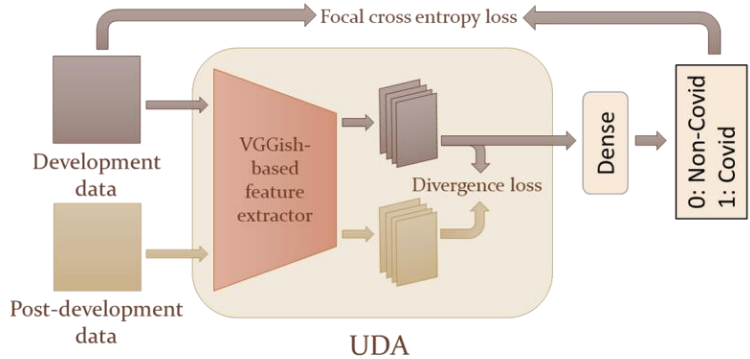

*Figure 5: Unsupervised domain adaptation (UDA) process. The model was fed with a batch of post-development data samples, along with a batch of samples from the development set and was then trained jointly (i) to correctly classify the labeled development data and (ii) to minimize the MMD distance between the embeddings of the development and post-development batches.*

During retraining, both loss functions were minimized simultaneously to enhance the model's adaptability to the post-development data while preserving its previous knowledge. Figure 5 shows an overview of the UDA method.

### Active learning

The second adaptation approach incorporated AL principles into the retraining process. Building upon the drift detection mechanism, a methodology was developed able to identify informative data points incorporating both diversity and uncertainty estimation [36,37]. Once a period of divergence was detected by the drift detection mechanism, uncertain instances were selected from the divergent batch of data. To achieve this, the z-scores of the model's outputs on the divergent data were calculated and the data samples whose output fell within one standard deviation around the mean value of the model's predictions were defined as uncertain. Samples within this uncertainty range were selected, thus prioritizing the inclusion of challenging and informative instances during retraining, with the ultimate goal to enhance the model's generalization capabilities.

Considering that this adaptation method involved selecting informative unlabeled data and utilizing them as labeled, it was essential to compare the results of AL with those obtained through random sampling. The number of randomly sampled data was equal to the number of data points employed in the adaptation phases of the AL approach.

## Results

### Baseline model

The baseline model was assessed in terms of its ability to accurately detect covid-19 cases in the presence of variations or shifts in the data. In particular, the performance evaluation on the two considered datasets for the development and post-development periods is reported in Figure 6.

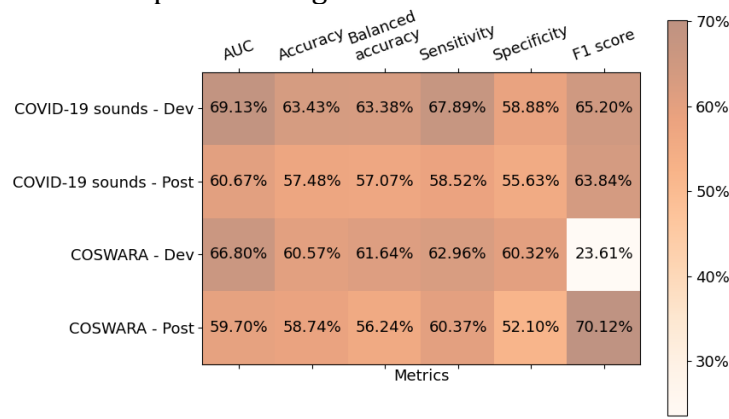

*Figure 6: Performance evaluation of the baseline model on the development and post-development data from the COVID-19 Sounds and COSWARA datasets. The model's discriminative ability was assessed in terms of the AUC, accuracy, balanced accuracy, sensitivity, specificity, and F1 score. The test subset sampled from the development data is referenced as Dev while post-development data (Post) refers to the entire post-development period.*

Based on the obtained results for the COVID-19 Sounds dataset, it is observed that the baseline model achieved superior performance on the test subset of the development period in terms of the Area Under the ROC Curve (AUC) (69.13%) and sensitivity (67.89%) compared to the best model performance reported in the literature [16] (66% AUC, 59% Sensitivity, 66% Specificity), despite considering a smaller amount of labeled data for training and validation (619 vs 1062 instances). The baseline model achieved a satisfactory F1 score (65.2%) but demonstrated quite low specificity, correctly classifying 58.88% of instances from the negative class. The performance of the baseline model on the post-development data demonstrated a significant decline in terms of the AUC, the F1 score, sensitivity, and specificity, as reported in Figure 6.

In the case of the COSWARA dataset, the baseline model displayed moderate discriminative ability on the development data, achieving an AUC value of 66.8%, while the accuracy, sensitivity, and specificity scores were equal to 60.57%, 62.96%, and 60.32%, respectively. The highly imbalanced distribution of the two classes in the development data (11.82% positive vs. 88.18% negative) posed a significant challenge for the model, as highlighted by the notably low F1 score, a metric that exclusively focuses on positive instances. The model's discriminative power on the post-development data presented a decline, as indicated by the AUC, specificity, and sensitivity, which are presented in Figure 6. The high value obtained for the F1 score metric was related to the presence of class imbalance with reversed minority (negative) and majority (positive) classes in the post-development data with respect to the class distribution of the development data, which led to a misleading perception that the model's performance had significantly improved. A more thorough analysis of this issue is provided in the Discussion section.

**Hyperparameters' tuning**

Tuning referred to hyperparameters related to data stream scanning and drift detection, which are summarized together with their examined values in Table 2. The tuning procedure considered hyperparameters used for data stream scanning (window length, overlap between successive windows, minimum batch size) and drift detection (kernel type and reference distribution for the MMD distance calculation and drift and threshold values of the CUSUM algorithm). The consideration of the minimum amount of data in a batch among the investigated hyperparameters enabled the inclusion of sufficient data during periods with a low data acquisition frequency, thus ensuring the robustness of the framework.

*Table 2: Investigated hyperparameters' values of the proposed framework.*

| Hyperparameters | | Values |
|---|---|---|
| Data stream scanning | Time window length | {7, 10, 14} days |
| | Time windows overlap | {70%, 60%, 50%, 40%, 30%, 20%, 10%, 0%} |
| | Minimum batch size | {0, 10, 20, 30, …, max batch size} |
| Drift detection mechanism | CUSUM Drift | {0.2, 0.3, 0.4, 0.5} |
| | CUSUM Threshold | {0.5, 0.6, 0.7, 0.8, 0.9, 1} |
| | Reference distribution | {Positive data, Negative data, All data} |
| | MMD kernel | {Linear, Polynomial, Gaussian} |

An offline nested approach, aiming at emulating the real-time operation scenario of the framework, was adopted for fine-tuning the framework's hyperparameters. In this context, a subset of the development data (70%), namely D-H set, was used for the creation of a baseline-H model. The D-H set was partitioned into training, validation, and test subsets based on a 60:20:20 split ratio. The rest of the development data (30%), named FT-H set, served for fine-tuning the hyperparameters' values (Figure 7).

The performance of the baseline-H model, measured in terms of the balanced accuracy on the test subset of the D-H set, was used as a benchmark against which the performance of the baseline-H model on the FT-H data batches was assessed. Any batch with performance below the benchmark was considered to be associated with a drift period and was properly labeled. The investigated hyperparameters' combinations were comparatively assessed in terms of their ability to correctly identify the labeled batches using appropriate metrics (accuracy, sensitivity, and specificity). The adopted offline nested approach is illustrated in Figure 8.

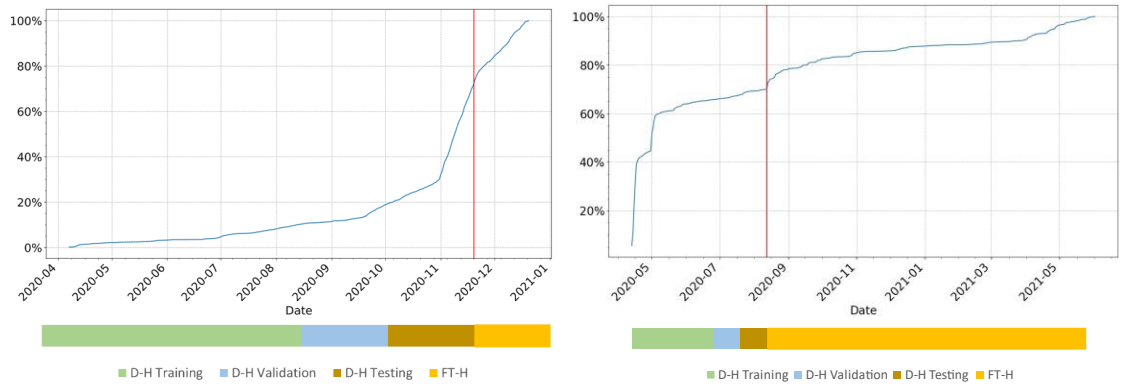

*Figure 7: COVID-19 Sounds (left panel) and COSWARA (right panel) data streams over time for the hyperparameters' fine-tuning. A 70:30 partition of the development data into D-H and FT-H sets (indicated by a vertical red line) was applied. The D-H set was further partitioned into a training, validation, and test subset using a 60:20:20 ratio.*

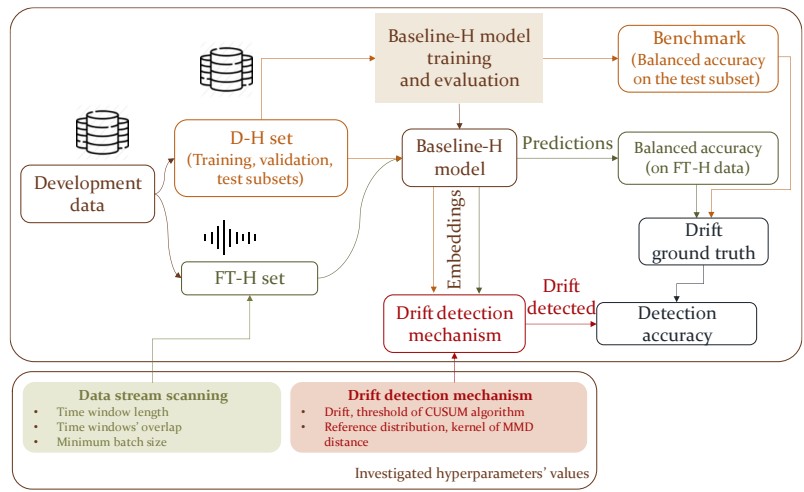

*Figure 8: Nested approach for the hyperparameters' fine-tuning. A 70:30 partition of the development data into D-H and FT-H sets was applied. The D-H set (further partitioned into a training, validation, and test subset using a 60:20:20 ratio) was used for developing a baseline-H model exclusively for hyperparameters' tuning. The balanced accuracy score on the test subset of the D-H set served as a benchmark. The FT-H set was used for fine-tuning the framework's hyperparameters to accurately detect batches where the model's performance fell below the benchmark.*

The balanced accuracy was selected for assessing model performance due to its ability to effectively capture the model's discriminative ability in the presence of class imbalance. Moreover, balanced accuracy was able to provide an appropriate measure even in the case of batches with a limited number of data instances, where the AUC score could not be applied.

*Table 3: The top ten hyperparameters' combinations in the case of the Covid-19 Sounds dataset. Detection accuracy, sensitivity, and specificity were used to evaluate the framework's ability to correctly identify the batches where the model's performance fell below the benchmark of the development period. The selected optimal combination is highlighted in bold.*

| Time window length (days) | Time windows overlap | Minimum batch size | CUSUM Drift | CUSUM Threshold | Reference distribution | MMD kernel | Accuracy | Sensitivity | Specificity |
|---|---|---|---|---|---|---|---|---|---|
| **7** | **60%** | **40** | **0.2** | **0.5** | **All data** | **Polynomial** | **0.80** | **0.66** | **1.0** |
| 7 | 40% | - | 0.4 | 0.8 | Positive | Polynomial | 0.75 | 0.33 | 1.0 |
| 7 | 40% | - | 0.3 | 0.7 | Positive | Gaussian | 0.75 | 0.33 | 1.0 |
| 7 | 40% | - | 0.3 | 0.8 | Positive | Gaussian | 0.75 | 0.33 | 1.0 |
| 7 | 40% | - | 0.4 | 0.5 | Positive | Gaussian | 0.75 | 0.33 | 1.0 |
| 7 | 40% | - | 0.4 | 0.6 | Positive | Gaussian | 0.75 | 0.33 | 1.0 |
| 7 | 40% | - | 0.4 | 0.7 | Positive | Gaussian | 0.75 | 0.33 | 1.0 |
| 7 | 40% | - | 0.5 | 0.5 | Positive | Gaussian | 0.75 | 0.33 | 1.0 |
| 7 | 40% | - | 0.5 | 0.6 | Positive | Gaussian | 0.75 | 0.33 | 1.0 |
| 7 | 40% | - | 0.2 | 0.5 | Positive | Polynomial | 0.75 | 0.33 | 1.0 |

Table 3 and Table 4 depict the top 10 hyperparameters' combinations based on the obtained accuracy for the COVID-19 Sounds and COSWARA cases, respectively. In the context of COVID-19 Sounds, the application of a 7-day window and a 3-day overlap between successive windows resulted in the highest accuracy score. The use of the total population of the development period as reference distribution and the Polynomial kernel for the MMD distance calculation along with a minimum batch size of 40 data samples led to the best performance. The CUSUM algorithm's drift and threshold values were set at 0.2 and 0.5, respectively, which corresponded to a tolerance range of 20% increase between successive batches and an accumulated increase of 50% beyond the tolerance range as optimal for generating alerts towards accurate detections.

In the case of COSWARA, the configuration including a 10-day window, no overlap between successive batches, and the total population of the development period as the reference distribution yielded the best results. In all top-performing combinations, a minimum of 20 data samples per batch was employed. Different combinations with variations of the CUSUM parameters and of the kernel used for the MMD distance calculation achieved similar optimal performance and triggered the same alerts. Based on this remark and in order to enhance the sensitivity of the CUSUM algorithm, the configuration utilizing the lowest values for the drift and threshold parameters, was selected.

Given that the proposed framework operates in a dynamically evolving manner, the performance evaluation of the drift detection mechanism during the post-development period is discussed in the next sections, along with the results obtained by applying the two different adaptation modules.

*Table 4: The top ten hyperparameters' combinations in the case of the COSWARA dataset. Detection accuracy, sensitivity, and specificity were used to evaluate the framework's ability to correctly identify the batches where the model's performance fell below the benchmark of the development period. The selected optimal combination is highlighted in bold.*

| Time window length (days) | Time windows overlap | Minimum batch size | CUSUM Drift | CUSUM Threshold | Reference distribution | MMD kernel | Accuracy | Sensitivity | Specificity |
|---|---|---|---|---|---|---|---|---|---|
| 10 | 0% | 20 | 0.3 | 0.7 | All data | Polynomial | 0.86 | 0.86 | 0.81 |
| 10 | 0% | 20 | 0.2 | 0.9 | All data | Linear | 0.86 | 0.86 | 0.81 |
| 10 | 0% | 20 | 0.3 | 0.9 | All data | Polynomial | 0.86 | 0.86 | 0.81 |
| **10** | **0%** | **20** | **0.2** | **0.7** | **All data** | **Linear** | **0.86** | **0.86** | **0.81** |
| 10 | 0% | 20 | 0.2 | 0.8 | All data | Linear | 0.86 | 0.86 | 0.81 |
| 10 | 0% | 20 | 0.3 | 0.8 | All data | Polynomial | 0.86 | 0.86 | 0.81 |
| 10 | 0% | 20 | 0.2 | 1.0 | All data | Linear | 0.86 | 0.86 | 0.81 |
| 7 | 0% | 20 | 0.2 | 0.5 | All data | Linear | 0.83 | 0.83 | 0.75 |
| 14 | 90% | 20 | 0.3 | 0.6 | All data | Polynomial | 0.82 | 0.81 | 0.75 |
| 14 | 90% | 20 | 0.2 | 0.5 | All data | Polynomial | 0.82 | 0.81 | 0.62 |

**Unsupervised domain adaptation**

A comparative assessment of the model's performance before and after each adaptation phase using the UDA approach was carried out. The performance of the baseline model, measured in terms of the balanced accuracy on the test subset of the development period, served as a benchmark for the evaluation of the model's performance after each adaptation period. For the COVID-19 Sounds dataset, the obtained balanced accuracy scores of the baseline model and the model after each adaptation phase are presented in Figure 9. It is evident that immediately after each alert period, the model's performance exhibited a considerable improvement, demonstrating the effectiveness of the proposed approach in mitigating the impact of concept drift. More specifically, following the first adaptation, the model showed an improved performance by up to 6% in terms of the balanced accuracy, which was maintained for four consequent batches (batches 2-5) before a second alert was triggered. Following the second adaptation, the model consistently outperformed the baseline model by up to 8%. After the third adaptation, the model enhanced its performance by more than 4%, remaining superior to the baseline model until the next drift detection, when the fourth adaptation led to a performance improvement of up to 15% which was maintained across ten batches (batches 19-28). After the fifth adaptation, the model exhibited a maximum performance increase of up to 24% and continued outperforming the baseline model until the end of the data stream. It is noteworthy that by correctly identifying periods of drift, the drift detection mechanism efficiently prevented the degradation of the model's performance in a timely manner while also contributing to sustaining the model's performance closer to the benchmark established during its development period.

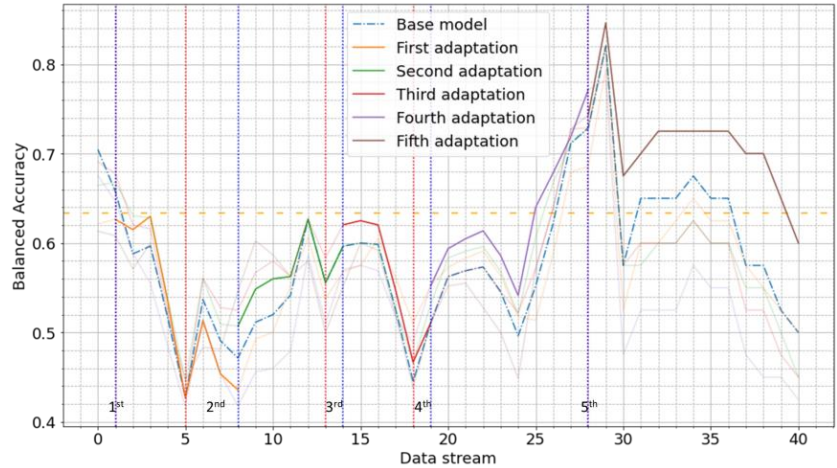

*Figure 9: The obtained balanced accuracy score across the data batches of the entire post-development period of the COVID-19 Sounds dataset. An orange dashed line is used to indicate the performance on the test subset of the development period (benchmark). Vertical red and blue dotted lines indicate the start and end of each alert period.*

In Figure 10, the discrimination performance of the adapted models is compared with that of the baseline model over their respective operational periods. It can be seen that after the first and fourth adaptation, each adapted model exhibited decreased performance in terms of the AUC, accuracy, sensitivity, and F1 score, but performed equally well or better than the baseline model in terms of the balanced accuracy, precision, and specificity. The other adaptations (second, third, fifth) resulted in a consistent and gradually increasing performance improvement according to all used evaluation metrics.

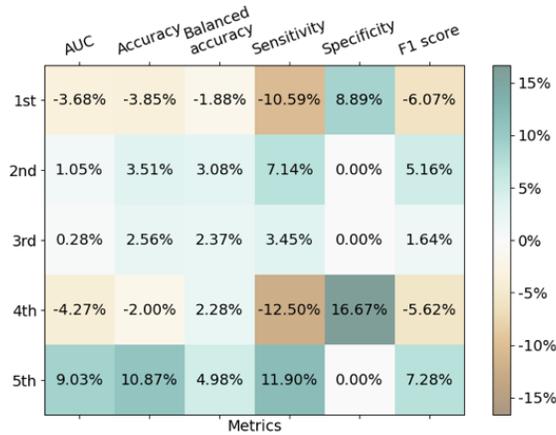

*Figure 10: UDA approach on COVID-19 Sounds dataset. Percentage change of model performance after each adaptation (1st – 5th) with respect to the baseline model.*

Regarding the COSWARA dataset, the obtained results are summarized in Figure 11. It is observed that the first adaptation led to a significant yet short-lasting decline in the model's performance, which persisted for two batches following the alert (until batch 4). Subsequently, the model exhibited a consistent and steady performance improvement, reaching up to a 20% increase compared to the baseline model's performance. Following the second adaptation, the model's performance presented a decline lasting for six batches after the alert, eventually returning to the baseline model's performance level. After the third adaptation, the model displayed a performance drop for one batch, followed by an increase of up to 10% that lasted for two batches. The fourth adaptation resulted in enhanced performance, demonstrating an improvement of up to 15% compared to the baseline model.

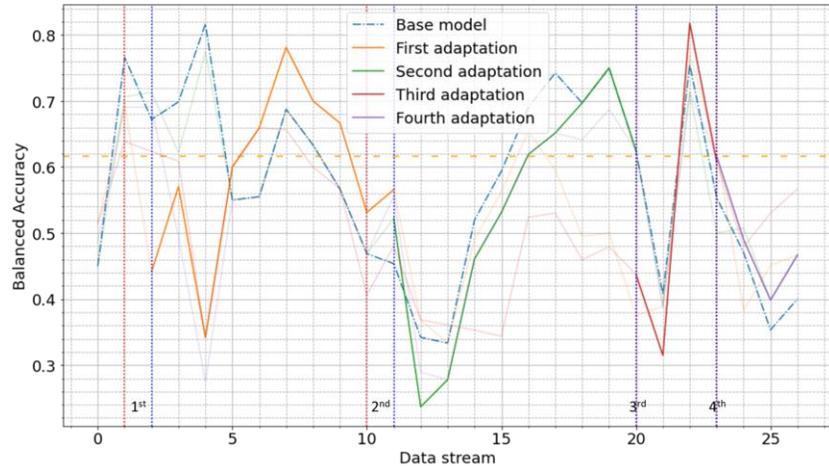

*Figure 11: Balanced accuracy score through the entire post-development period on COSWARA dataset. An orange dashed line is used to indicate the performance on the test subset of the development period. Vertical red and blue dotted lines indicate the start and end of each alert period.*

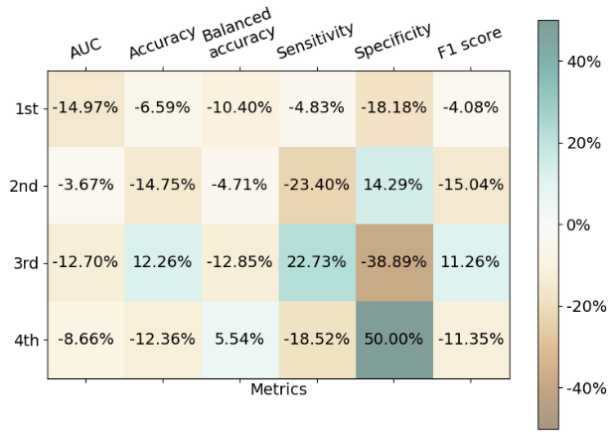

*Figure 12: UDA approach on COSWARA dataset. Percentage change of model performance after each adaptation (1st – 4th) with respect to the baseline model.*

Although in the case of the COSWARA dataset, the UDA method was particularly challenged in maintaining the model's performance, the drift detection mechanism consistently demonstrated effectiveness by generating timely alerts related to anticipated declines in the model's performance. Notably, the first alert was triggered during a phase where the model's performance was gradually decreasing, although it still remained above the benchmark. The second alert was triggered immediately after a performance drop below the benchmark level, preventing an even steeper decline. Similarly, the third and fourth alerts were produced amid significant performance deterioration, aligning precisely with instances where the model fell below the benchmark level, thus indicating the robustness of the drift detection mechanism. The comparative assessment of the performance of the baseline model and the model after each adaptation (Figure 12) demonstrated that the implemented adaptations resulted in varying performance in terms of the different evaluation metrics.

**Active learning**

The proposed AL approach was evaluated by comparing the performance of the model after each AL-based retraining phase with that of the baseline model as well as the model following retraining using random sampling. Looking at the COVID-19 Sounds dataset, Figure 13 demonstrates the observed balanced accuracy score across the entire data stream, indicating a substantial and lasting improvement with respect to the baseline model following each adaptation. After the first adaptation, the model exhibited improved performance by up to 20% until the second alert, while a similar improvement was achieved following the second adaptation. After the third adaptation, the model's performance demonstrated a significant increase of up to 30%, sustained over a broad period of 15 batches. After the fourth adaptation, the model showcased outstanding performance, surpassing a 95% balanced accuracy score while achieving an improvement of up to 25% compared to the baseline model.

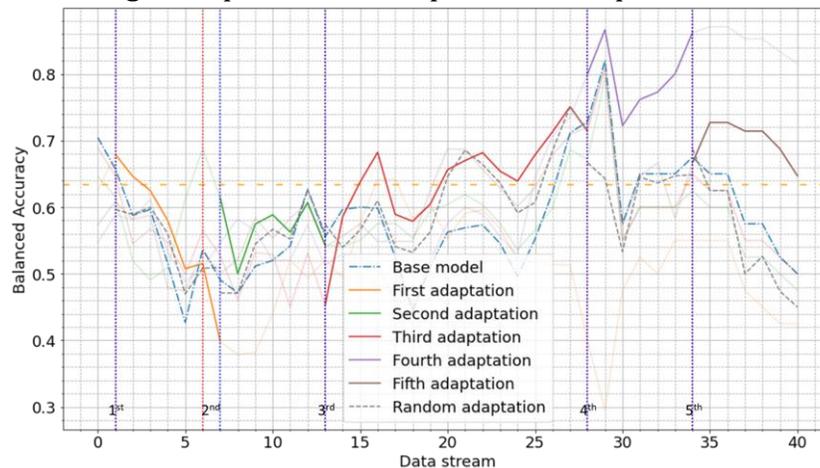

*Figure 13: Balanced accuracy score through the entire post-development period on COVID-19 Sounds dataset. An orange dashed line indicates the performance on the test subset of the development period. Vertical red and blue dotted lines indicate the start and end of each alert period.*

The performance comparison with the random sampling approach highlighted the superiority of the proposed AL approach, indicating its ability to identify informative data. Specifically, the random sampling approach outperformed AL only in four batches (i.e., batches 7, 12, 13, and 21) out of the 41 batches of the data stream. AL significantly outperformed the random sampling approach in all other cases. Similar conclusions can be drawn from Figure 14, where it can be seen that the majority of adaptations based on AL yielded increased performance in terms of all the considered evaluation metrics with respect to the baseline model and the random sampling approach.

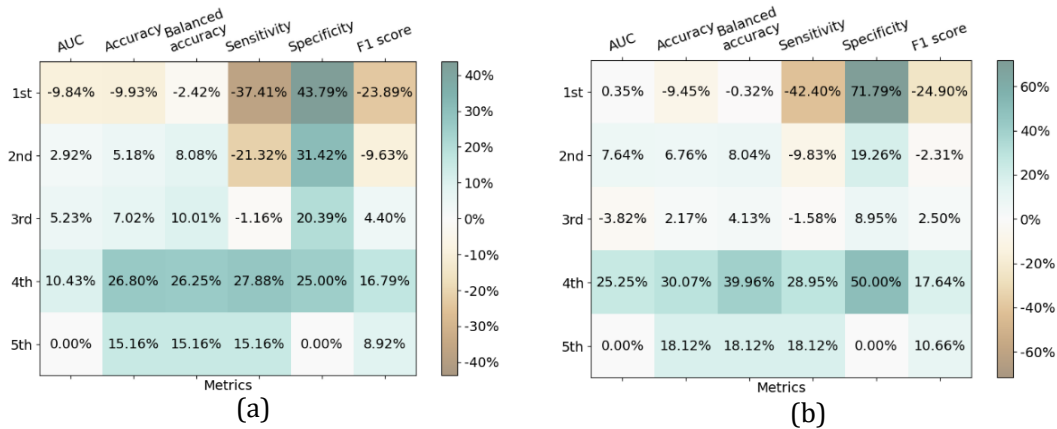

(a)          (b)

*Figure 14: AL approach on the COVID-19 Sounds dataset. Percentage change of model performance after each adaptation (1st – 5th) with respect to the baseline model (left panel) and the model after retraining through random sampling (right panel).*

In the case of the COSWARA dataset, Figure 15 shows that model adaptations led to improved performance during the post-alert periods. Following the first adaptation, there was a marginal increase in the balanced accuracy score, with the model slightly outperforming the baseline model until the subsequent alert. After the second adaptation, a notable enhancement of up to 20% in the balanced accuracy score was observed in the subsequent batches, while the fourth adaptation resulted in performance improvement of up to 25%. In terms of the third and fifth adaptations, although the model's performance exhibited significant enhancement, reaching up to 40% and 60%, respectively, performance fluctuations were also observed, with the model being outperformed by the baseline model in some batches.

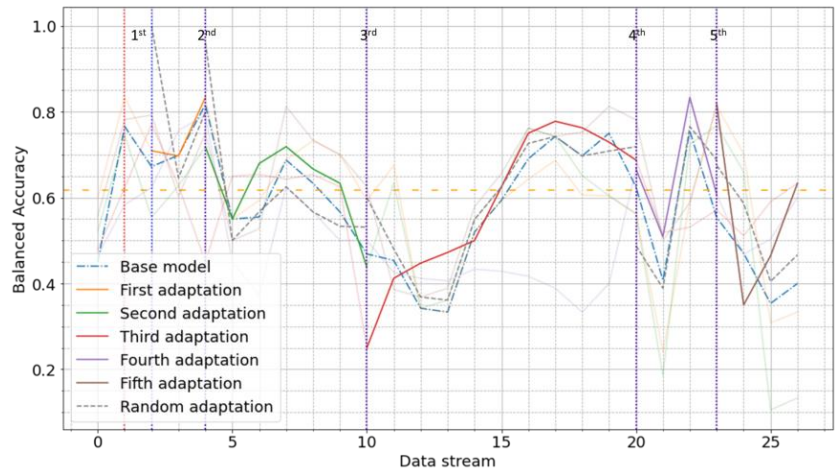

*Figure 15: Balanced accuracy score through the entire post-development period on COSWARA dataset. Orange dashed line indicates the performance on the test subset of the development period. Vertical red and blue dotted lines indicate the start and the end of each alert period.*

When comparing the performance of the AL method with the results obtained through random sampling, it becomes evident that the former method demonstrated superior performance throughout the majority of the data batches except for specific instances (within batches 2, 4, 10, 11, 14, 24). This highlighted AL's efficiency in identifying informative data.

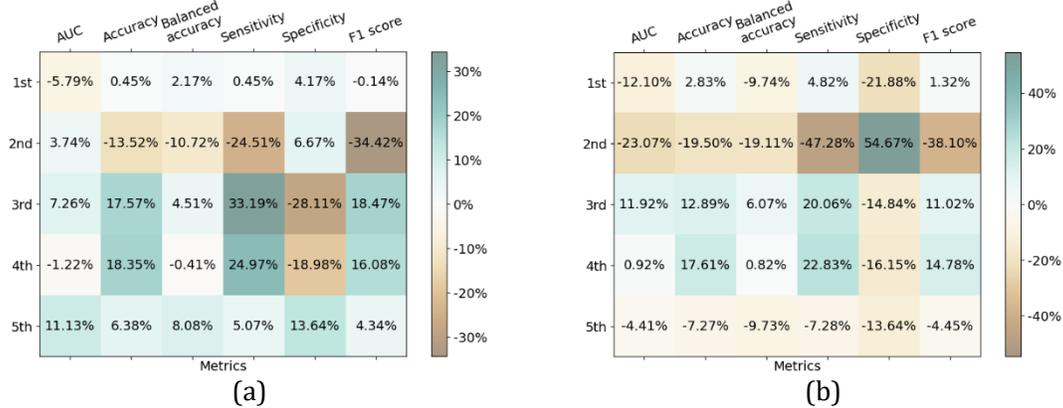

*Figure 16: AL approach on the COSWARA dataset. Percentage change of model performance after each adaptation (1st – 5th) with respect to the baseline model (left panel) and the model after retraining through random sampling (right panel).*

Further insights on the model's performance after each adaptation, with respect to the baseline model and the model after retraining based on random sampling, can be drawn from Figure 16. It can be seen that the AL approach led to an overall improvement in the model's performance over the baseline model, except for certain metrics/adaptations (Figure 16(a)). This is evident in the second adaptation, where the model exhibited an improved AUC score and specificity, but inferior performance in terms of the remaining evaluation metrics. The comparison with the random sampling approach (Figure 16(b)) revealed a more intricate pattern. The third and fourth adaptations, spanning for more than half of the entire data stream, clearly had a positive impact on the model's performance. However, the first adaptation resulted in higher sensitivity levels at the cost of lower specificity, while the second and fifth adaptations yielded lower performance.

### Discussion

The evaluation of the proposed framework for detecting data drift and performing model adaptation provided evidence regarding the framework's ability to maintain model performance, thus highlighting its potential to facilitate the identification of new cases in a dynamic, non-stationary environment caused by a pandemic.

A baseline model able to detect covid-19 positive cases using cough recordings was trained and evaluated. During the development period, the model achieved an AUC of 69.1% and 66.8% on the COVID-19 Sounds and COSWARA datasets, respectively. However, in the post-development period, there was a notable decline in the baseline model's performance, reflected to an AUC of 60.7% and 59.7%, respectively, thus suggesting the potential presence of concept drift. These findings motivated the development of the proposed approach, which leveraged robust drift detection and efficient adaptation mechanisms to maintain the model's performance in the face of evolving data distributions.

The obtained results indicated the efficacy of the proposed drift detection mechanism and provided evidence regarding its ability to enhance the robustness and adaptability of deep learning models in dynamic environments. The combination of the MMD distance monitoring and the use of the CUSUM algorithm for adaptive detection of abrupt changes, which reflect a growing divergence between the reference distribution (development data) and the post-development data, enabled the timely and robust detection of performance degradation. The use of the CUSUM algorithm, tailored to the characteristics of each dataset, ensured the generation of accurate alerts for significant changes in the monitored MMD distance, thus minimizing false alerts and preventing unnecessary interventions.

Two distinct retraining strategies based on UDA and AL were comparatively assessed in terms of their ability to mitigate performance degradation. The UDA approach enhanced the model's ability to learn from unlabeled data, while AL facilitated the selection of informative instances for targeted retraining. The results obtained from the analysis of the COVID-19 Sounds dataset showed that the UDA approach significantly improved the model's discriminative ability. The comprehensive examination of the subsequent adaptation phases based on multiple evaluation metrics revealed deviations from the baseline model's performance, most of which corresponded to an improvement in the balanced accuracy, ranging from 10% to 20%.

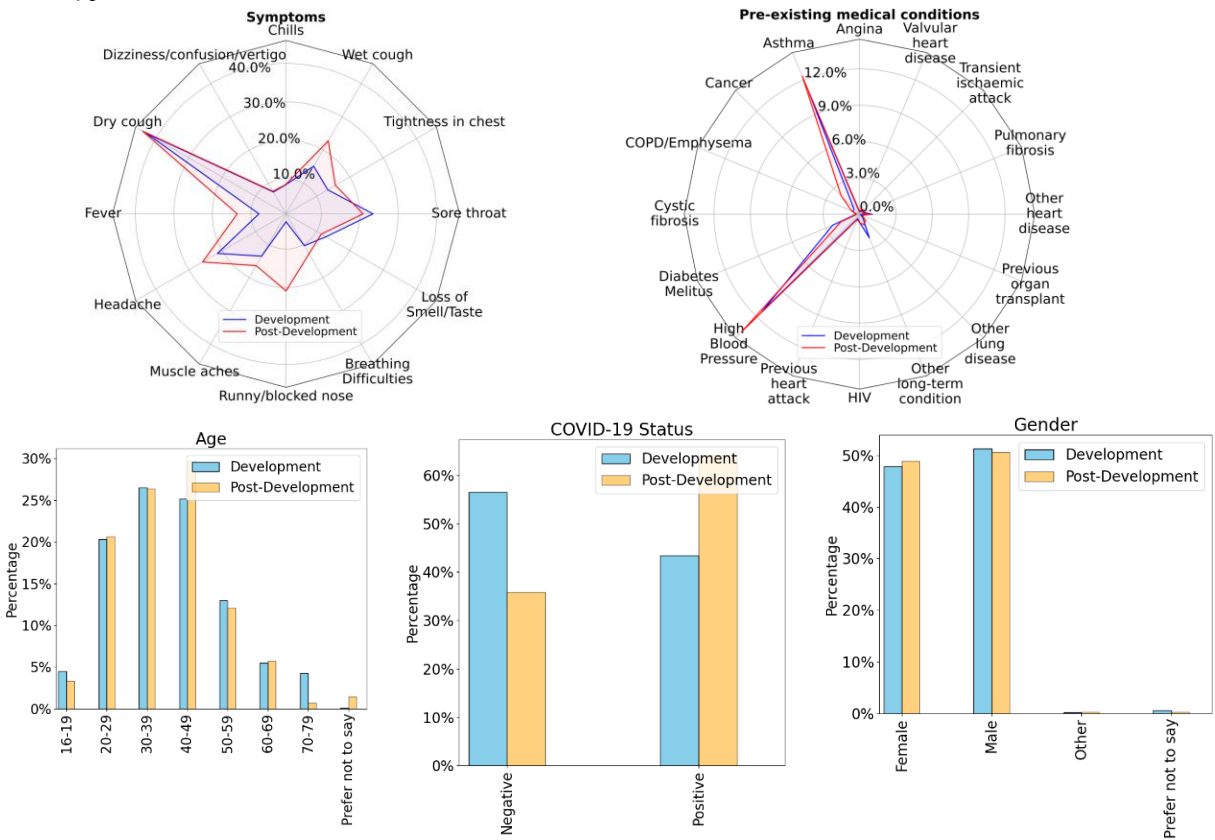

*Figure 17: Descriptive statistics for COVID-19 Sounds development and post-development data reveal moderate changes. The disease exhibited moderate shifts in both its prevalence and the frequency of related symptoms. The two data subsets shared similar characteristics in terms of age, gender, and medical history of individuals.*

The above findings align with UDA's core advantages, which include cost-effectiveness and adaptability to dynamic environments through the model's adaptation to the target domain's data distribution without requiring labeled target domain samples. This approach is particularly valuable when labeled data from the target domain are scarce or expensive to obtain, as is often the case in emerging pandemic scenarios.

The application of the UDA approach on the COSWARA dataset yielded less consistent results. The overall comparison between the adapted models and the baseline model revealed that the adaptation had diverse effects on the model's performance in terms of the considered evaluation metrics.

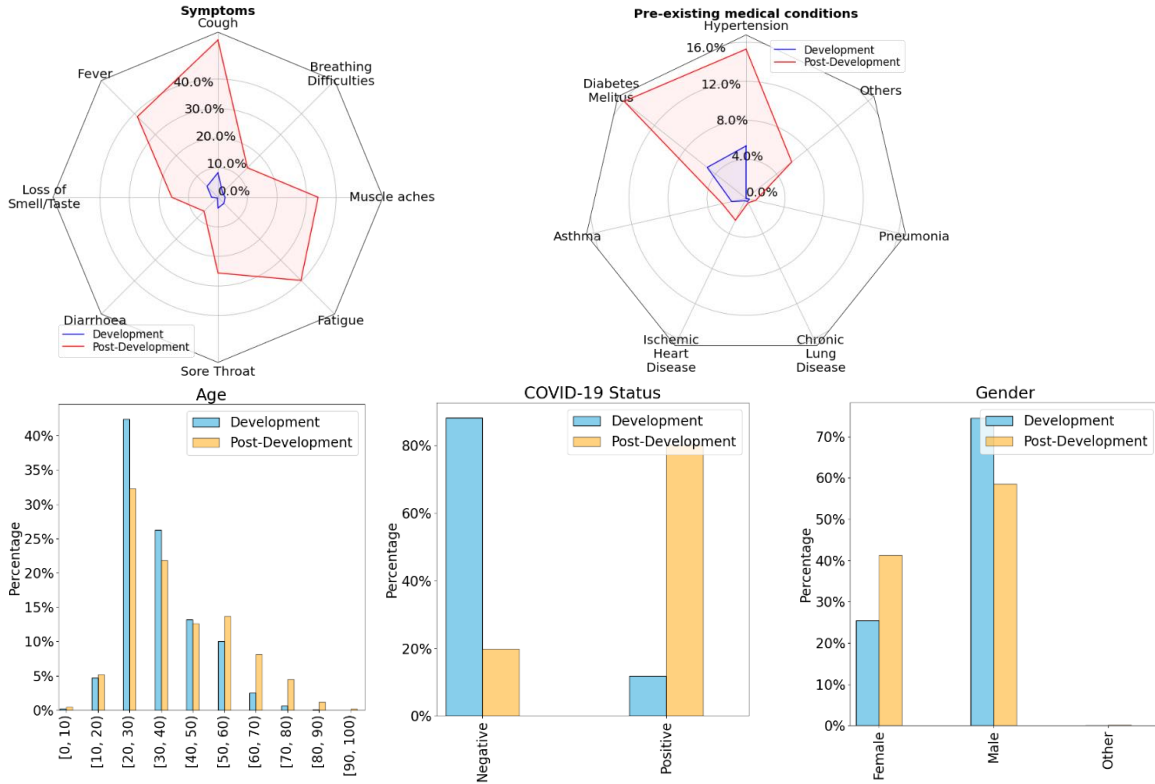

*Figure 18: Descriptive statistics for COSWARA development and post-development data reveal profound differences in terms of demographic characteristics, symptoms, and pre-existing medical conditions between the development and post-development period. The representation of positive and negative classes in the development data is reversed in the post-development data.*

The difference in the effectiveness of the UDA approach on the COVID-19 Sounds and COSWARA data may be attributed to differences in the datasets' characteristics. Figure 17 and Figure 18 illustrate selected descriptive statistics of the development and post-development data of the COVID-19 Sounds and COSWARA dataset. As per [16], the COVID-19 Sounds dataset used in the present study had underwent meticulous curation to eliminate biases as a result of methodological decisions, thus enabling the development of unbiased models. In the case of COSWARA dataset, significant differences were observed in terms of covid-19 prevalence and the frequency of related symptoms between the development and post-development data, which may be attributed to the presence of age and gender biases [59]. In the present study, handling the data in chronological order implicated different levels of data biases present across the various adaptation periods, which may arise in emerging pandemic scenarios.

The AL approach resulted in more prominent improvement in the models' performance compared to UDA. This underscored the power of actively selecting informative samples for labeling, which aids in refining the model's understanding of the target domain. Thus, by optimizing both adaptation to the target data and use of labeling resources, AL is considered promising for ensuring model performance in data scarce scenarios, such as during a pandemic.

Given that both UDA and AL achieved varying levels of performance improvement on both datasets, it is essential to consider their limitations and potential challenges. UDA relies on the assumption that the source and target domains share some underlying similarity. In the presence of significant differences, adaptation might not yield substantial improvements. On the other hand, AL's performance is determined by human labeling expertise, which is associated with the rise of the related costs and depends on the reliability of the existing diagnostic tests. If the chosen samples are mislabeled, the model's performance could suffer. Moreover, AL's performance is sensitive to the selection of labeled samples, which might introduce biases.

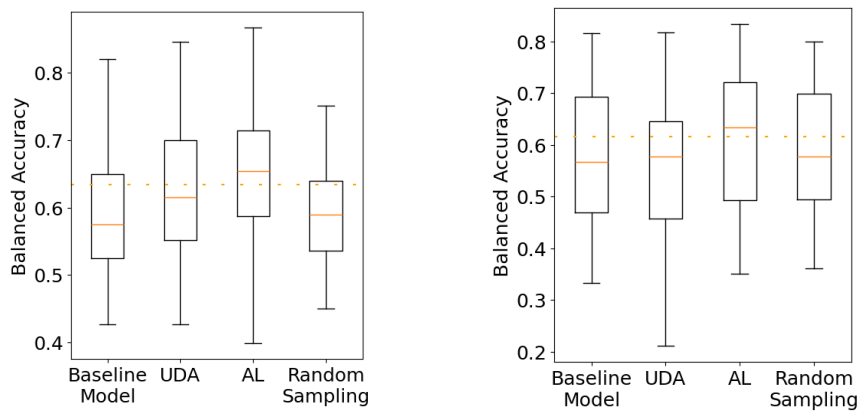

*Figure 19: Box plots of the balanced accuracy scores across the entire post-development period using the baseline model, the UDA approach, the AL approach, and the random sampling approach for the COVID-19 Sounds (left) and COSWARA (right) datasets. The orange dashed line indicates the performance of the baseline model on the test subset of the development period (benchmark).*

The above observations suggest that the proposed adaptation mechanisms effectively addressed the individual challenges linked to the special characteristics of each dataset and mitigated the effects of concept drift during critical periods corresponding to batches in proximity to the alert periods. Figure 19 summarizes the model's performance obtained by applying each adaptation approach on the post-development period of the two datasets and shows that both approaches succeeded in maintaining the models' performance closer to the development period's benchmark. These results highlight the importance of combining effective drift detection mechanisms and intelligent adaptation modules in addressing concept drift. Together, these components form a robust framework that enables the model to continuously adapt to changing data conditions, thereby maintaining its discriminative power and overall performance over time.

The significance of the proposed approach lies in its reliance on unsupervised techniques. By minimizing the dependence on labeled data, the proposed framework enables the accurate detection of covid-19 cases even in the absence of comprehensive labeling resources. This aspect becomes particularly crucial when considering the value of a deep learning-based detection model during the early stages of a new pandemic or when dealing with emerging viral variants that may not be adequately detected by existing diagnostic tools. Thus, the proposed framework is able to contribute towards a more generalizable approach that can be applied to future pandemics or novel variants. By collecting knowledge and formulating a well-defined framework, a basis for rapid adaptation and deployment of disease detection tools is established, ensuring timely and accurate identification of infectious diseases.

## Acknowledgments

This work was supported within the framework of the Smarty4covid project, which is funded by the Hellenic Foundation for Research and Innovation (project 5020).

## Conflicts of Interest

There are no conflicts of interest to disclose.

## Data Availability

The COSWARA dataset used in the current study is publicly available in the GitHub repository and can be accessed through the link https://github.com/iiscleap/Coswara-Data. Access to the COVID-19 Sounds dataset was granted by the Department of Computer Science, Cambridge University, but restrictions apply to the availability of these data, which are not publicly available and were used under license for the current study. Data can be made available to academic institutions for academic research purposes upon the submission of relevant request to the Mobile Systems group and the signing of a Data Transfer Agreement. Please contact covid-19-sounds@cl.cam.ac.uk to obtain it.

**Abbreviations**
UDA: unsupervised domain adaptation
AL: active learning
MMD: maximum mean discrepancy
CNN: convolutional neural network

PCR: polymerase chain reaction
CUSUM: cumulative sum
AUC: area under the ROC curve